\documentclass[pre,superscriptaddress,twocolumn,a4paper,showpacs]{revtex4-1}
\usepackage{graphics}
\usepackage{amssymb}
\usepackage{wasysym}
\usepackage{latexsym}
\usepackage{graphicx}
\usepackage{epsfig}
\usepackage{subfigure}
\usepackage{todonotes}
\usepackage{soul}

\begin{document}

\title{The  paradox of productivity during quarantine: an agent-based simulation}

\author{Peter Hardy}
\affiliation{%
School of Computing and Communications,
  Lancaster University,
  Lancaster, Lancashire, LA1 4WA, UK
  }%
\author{Leandro Soriano Marcolino}%
\affiliation{%
School of Computing and Communications,
  Lancaster University,
  Lancaster, Lancashire, LA1 4WA, UK
  }%

\author{Jos\'e F.  Fontanari}
\affiliation{%
Instituto de F\'{\i}sica de S\~ao Carlos,
  Universidade de S\~ao Paulo,
  Caixa Postal 369, 13560-970 S\~ao Carlos, S\~ao Paulo, Brazil
  }%


\begin{abstract}
 Economies across the globe were brought to their knees due to lockdowns and social restriction measures to contain the spread of the SARS-CoV-2, despite the quick switch  to remote working. This downfall may be partially  explained  by  the ``water cooler effect'', which holds that higher levels of social interaction  lead to higher productivity due to a boost in people's mood.  Somewhat paradoxically, however, there are reports of increased productivity  in the  remote working  scenario. Here we address quantitatively this  issue using a variety of experimental  findings of social psychology that address the interplay between mood,  social interaction and productivity to set forth 
an agent-based model  for  a workplace  composed of  extrovert and introvert agent stereotypes  that differ solely on their propensities to initiate a social interaction. We find  that the effects  of curtailing social interactions depend on the proportion of the  stereotypes in the working group:  while the social restriction  measures  always have a  negative impact on the productivity of groups composed predominantly of introverts, they may actually improve the productivity of  groups composed predominantly of extroverts. Our results offer a proof of concept that the paradox of productivity during quarantine can be explained by taking into account the distinct effects of the social distancing  measures on extroverts and introverts.
\end{abstract}

\maketitle


\textbf{Introduction.} -- The interest on the trade-off  between  work productivity and social interaction  is  not a present-day fad,   as attested by this verse of the 19th century writer Maria Edgeworth:  ``All work and no play makes Jack a dull boy,
All play and no work makes Jack a mere toy" \cite{Edgeworth_25}. In fact, this is a staple subject of social psychology  that addresses the influence of loneliness or, more generally,  of mood  on  human cognitive function, often with discrepant   findings (see, e.g., \cite{Cacioppo_09,Stumm_18}). Recent  lockdowns and social restriction measures  that isolated workers from their peers have  placed these issues at the forefront of public attention, with news outlets reminding their customers that they do not need to be productive, and to expect a reduction in overall productivity during this time.  But why is it that productivity would decrease during a period that one would think people have more time on their hands than ever before? 

Typical answers to this question are in line with the so-called   ``water cooler effect'' that has been highlighted (and satirized) in how effective placing an inanimate object for people to congregate around can stir up casual conversations, with many psychologists believing that this can also increase company productivity \cite{Pentland_09}.  However, this is not an obvious conclusion since  for a singular worker the extra down time communicating  may actually detract from its productivity as a whole. 
The economic literature offers more solid explanations for the positive correlation  between  social interaction or communication among co-workers and productivity \cite{Cornelissen_17}. For instance, the peer pressure that generates feelings of guilt or shame when one's performance falls short of the expectation may increase  focus on the work  and hence increase productivity \cite{Kandel_92,Bandiera_10}. In addition,  it is through social interactions that coworkers learn from each other and build up skills that they otherwise would not have. This  so-called knowledge spillover  is the usual explanation for the  strong link between economic growth and the concentration of people
 in cities \cite{Feldman_99,Bettencourt_07}.

Here we address quantitatively the  productivity and social interaction issue using an   agent-based model to simulate a workplace  
scenario where the agents exhibit two social stereotypes, viz., extroverts and introverts, that differ solely on their propensities to initiate a  conversation. This is in accord with the well-established view that  extroverts and introverts equally enjoy engaging in social interactions and in other pleasurable activities,  but differ in the likelihood that they would be the instigators of those activities \cite{Feist_12}.
Social distancing is modeled by controlling  the number of attempts an agent makes to find a conversation partner.  The  motivation to work (mood) is assumed to increase  with the time spent  talking and decrease with the time spent alone.   Moreover, the instantaneous productivity of a lone agent increases  with its motivation to work. 
These assumptions are grounded on the unsurprising findings that positive affect (i.e.,  the propensity to be in a good mood) increases significantly after social interaction \cite{Phillips_67,McIntyre_91},  facilitates creative problem solving \cite{Isen_87}  and boosts productivity in general \cite{Ellis_88,Oswald_15}.

We find that the effect of curtailing social interactions depends on the proportion of the different stereotypes in the group. For instance, while
social restriction measures always have a  negative impact on the productivity of groups composed predominantly of introverts, they  may actually improve the productivity of  groups composed predominantly of extroverts. In addition, within a same working group  the productivity of the two stereotypes is affected differently by those  measures. These results may explain the  paradoxical findings  that  productivity is increased  in some  remote work experiments \cite{Bloom_15}.


\textbf{Model.} -- We model the dynamics of a group of $N$ agents that interact socially and work  during a 480-minute  (eight-hour) workday. We first  make the fair assumption that for all agents they are more productive when they are motivated (i.e., in a good mood to work) \cite{Isen_87,Ellis_88,Oswald_15}, and their motivations increase when they  participated  in some form of social interaction \cite{Phillips_67,McIntyre_91}.  In this  framework,  motivation has both an emotional dimension, in the sense that talking with peers improves the mood of the agent and helps it to focus on the task, and an informational dimension, in that  talking with peers may result in the acquisition of  valuable information that can help the agent to consummate  its task,  which is the  knowledge spillover mechanism mentioned previously.

We assume that the motivation of agent  $k=1, \ldots, N$  is determined  by the integer parameter $L_k = 1,2,\ldots, L_{max}$  and that at the beginning of the day,  the   agents' motivations  are the lowest, i.e., $L_k = 1$.  In addition, we assume that the instantaneous productivity of an agent at time $t=0, 1, \ldots, T$ is  $P_k (t) =  L_k -1 $ if agent $k$ is not talking and $P_k (t) = 0$ if agent $k$ is talking. 
This is in line with the experimental findings that  being with other people would be associated with poorer cognitive task performance than when being alone \cite{Stumm_18}. 
Since  we measure time in minutes, we set $T=480$. Hence, at time $t=0$  the instantaneous productivity is zero for all agents.  

 The only way an agent can increase its motivation and, consequently, its productivity, is by engaging  in social interaction.
 We  introduce two different social stereotypes our agents can be, viz., extroverts and introverts, which determine their propensities  to seek and engage in social interaction. In particular,  if agent $k$  has stereotype $l$, where $l=e$ for extroverts   and  $l=i$ for introverts,  the probability that it instigates a conversation  is
 \begin{equation}\label{prob_e}
 p_k^l = \left ( T/\tau^l - L_k \right )/\left ( T/\tau^l - 1 \right )
 \end{equation}
 for $ L_k \leq T/\tau^l $ and $ p_k^l = 0$, otherwise. Here $\tau^e$ and $\tau^i$ are parameters measured in minutes that are necessary to make eq.\ (\ref{prob_e}) dimensionally correct. Throughout this paper we set $\tau^e =1$ without loss of generality, but set $\tau^i > 1$ in order to guarantee that $ p_k^e \geq  p_k^i$, i.e., that  the extroverts are more likely to engage in social interaction than the introverts, 
in agreement with the personality theory  findings \cite{Feist_12}. In addition, $ p_k^e =  p_k^i = 1$ for  $L_k =1$  so that when the motivation of agent $k$ is at the bottom line, it will try to engage in social interaction with certainty.  The propensity to  instigate a conversation decreases linearly with the motivation to work until it vanishes altogether when the  motivation parameter reaches  a threshold value that differs for the two social stereotypes. This decrease is justified  because  no work is done while the agent is talking  and so we should expect that  the higher the agent's  motivation to work is, the lower its propensity to socialize. The different threshold values  are necessary to account for the existence of two unambiguous stereotypes. 

Regardless of its stereotype, once a target agent decides to instigate a conversation, it selects  a number $m$ of  contact attempts, where $m =0, 1, \ldots$ is a random variable drawn from a Poisson distribution of parameter $q$.  In each contact attempt, a  peer is selected at random among the $N-1$ agents in the group and, in case the selected agent is not in a conversation at that time,  a conversation  is initiated   and the target agent halts its search for a partner.   The duration of the conversation $d$ is given by  a random integer selected uniformly in $\{1, ..., D\}$. We note that  a conversation involves  two agents only and  the agent that is approached by the target agent is obliged to accept the interaction, regardless of  its motivation and stereotype. This pro-social behavior is chosen in order to not further complicate the model, but it can be justified in terms of workplace social norms \cite{Johns_05}.

To complete  the model  we need to give a prescription for changing the motivation parameter $L_k$ with $k=1, \ldots, N$, which is the central  factor in the determination of  the social behavior of the  agents.  We assume simply that the motivation of  agent $k$  increases by one unit for each minute it spends in a conversation
and decreases by one unit for each minute it spends working alone. This is in accordance with the previously mentioned experimental findings that mood improves after social interaction  \cite{Phillips_67,McIntyre_91} and that  being alone rather than with others while doing a task results  in worse mood but better cognitive task performance \cite{Stumm_18}.  Moreover, we set $L_k = 1$  as the lower bound of  the motivation  parameter so that a lone agent $k$ at the motivation bottom line  $L_k = 1$  cannot have its motivation reduced any further. 

We note that our prescription to change the motivation of the agents implies that  the motivation  parameter is bounded from above by $L_{max} =480$, so that the probability that an extroverted agent instigates a conversation, eq.\  (\ref{prob_e}), is never zero during the eight-hour workday. 
This produces a qualitative distinction between the two stereotypes since the extroverts are, in principle,  always willing to initiate a social interaction (i.e., $p_k^e > 0$), whereas the introverts will not instigate a conversation if their motivation to work is above the threshold $T/\tau^i$.  Throughout this paper  we set $\tau^i = 5$ but note that this choice is immaterial,  provided that $T/\tau^i$ is sufficiently distinct from the upper bound $T/\tau^e$ to justify the  existence of two stereotypes. 

In summary,  we implement the synchronous or parallel update of the $N$ agents as follows. At $t=0$ we set $L_k = 1$ and $P_k =0$ for all agents $k=1, \ldots, N$. The update procedure at time $t$ begins with the  selection of a random order of update of the  $N$ agents,  so that at the end of the procedure  all agents are updated and we can increment the  time  from $t$ to $t+1$.  Then we check the status of the agent to be updated -- the target agent --  to determine if it is participating in a conversation or not.  In case it is, we increment its motivation parameter by one unit. In case it is not, we decrement its motivation parameter by one unit and test  whether it  initiates 
 a conversation or not using the probabilities   given in eq.\  (\ref{prob_e})   in accord with the stereotype of the target agent.
 The simulation ends at $t=T$. 

\textbf{Results.} -- We consider  working  groups composed of $N^e$  extroverts and $N^i = N - N^e$ introverts and focus mainly on  the mean cumulative productivity of the stereotypes at time $t=1,\ldots, T$,
 \begin{equation}\label{Pi_l}
 \Pi^l (t)  = \frac{1}{N^l} \sum_{k \in \mathcal{S}^l} \pi_k (t) 
 \end{equation}
 where
 \begin{equation}\label{pi_k}
 \pi_k (t)  = \frac{1}{t}  \sum_{t'=0}^t  P_k (t')
 \end{equation}
and the superscript   $l = e, i$ specifies the social stereotypes of the agents, as before.  Here the sum over $k$ is restricted to the subset 
$ \mathcal{S}^l$ of agents  with stereotype $l$, whose cardinality is $N^l$. 
We recall that $ P_k (0) =0$ for all $k$.  The mean cumulative productivity of the whole  group is  simply $\Pi^w  =  \left ( N^e \Pi^e +  N^i \Pi^i \right) /N$, where we have omitted the dependence on $t$ for simplicity.  It is also of interest to know the mean motivation of the stereotype subgroups,  
 \begin{equation}\label{Pi_l}
 \Lambda^l (t)  = \frac{1}{N^l} \sum_{k \in \mathcal{S}^l} L_k (t) 
 \end{equation}
 with $l = e, i$ as before. Then the mean motivation of the whole group is  $\Lambda^w  =  \left ( N^e \Lambda^e +  N^i \Lambda^i \right) /N$.

Our main interest  here is  $\Pi^l (t) $ and $\Lambda^l (t)$  evaluated  at the end of the day, i.e, at $t=T$. These quantities are averaged over   $10^3$ independent runs for each setting of the model parameters. In particular,  in the following we vary the fraction  of extroverts $\eta = N^e/N $ and  the mean number of contact attempts  $q$, but  fix the group size to $N=100$ and the maximum duration of a conversation to $D=20$. As already pointed out, we set $\tau^e = 1$ and $\tau^i = 5$.

\begin{figure}[t] 
\centering
 \includegraphics[width=1.0\columnwidth]{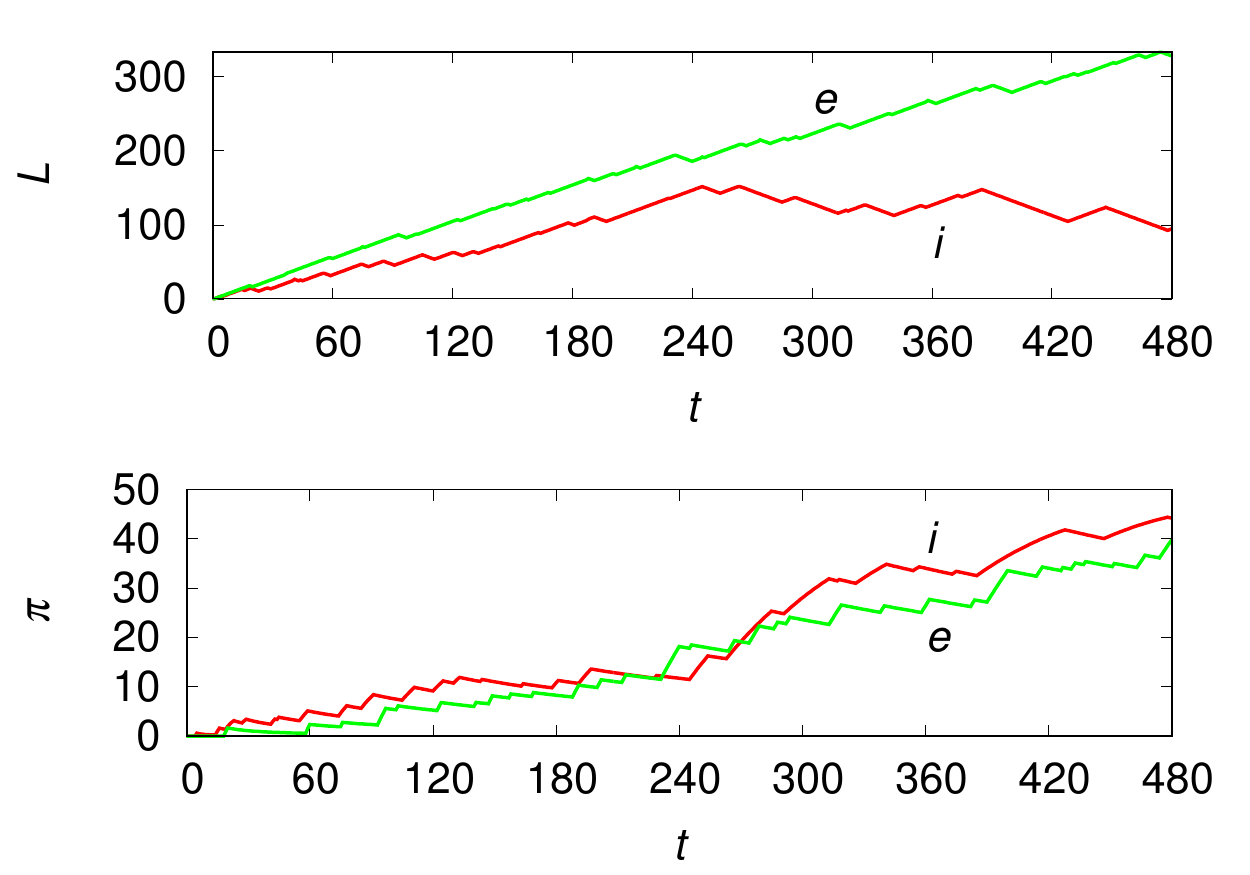}  
\caption{Motivation  (upper panel) and mean cumulative productivity  (lower panel) of  typical extrovert  ($e$) and  introvert  ($i$) agents for a group of size $N=100$ with equal ratio of extroverts and introverts, i.e., $\eta = 0.5$. The maximum duration of a conversation is $D=20$ and the mean number of contact attempts  is $q=2$. 
 }  
\label{fig:1}  
\end{figure}

 To better appreciate the  assumptions of the model,  fig.\  \ref{fig:1} shows  typical  time dependences of the  motivation and   mean cumulative productivity of two agents with distinct stereotypes for a single run.   The time intervals where the mean cumulative productivity  decreases correspond to the periods when the agent is participating in a social interaction and are associated with the increase of its motivation. We note that, whereas the cumulative  productivity is a non-decreasing function of $t$, the mean cumulative  productivity, eq.\ (\ref{pi_k}),  decreases with increasing $t$ in the time intervals where the instantaneous productivity of  the agent is zero. In addition, the motivation decreases in the periods where the productivity increases. Most of the time in this run,  the introvert agent exhibits a higher mean productivity than the extrovert agent, despite  its  lower motivation. In fact, the motivation parameter of the introvert stabilizes and fluctuates around the value $L = T/\tau^i = 96$, whereas the motivation parameter of the extrovert shows a tendency to increase linearly with $t$ on  average.
It is easy to understand the behavior pattern of the introvert's motivation if we note that  an introvert agent with $L \geq T/\tau^i $ will not attempt to instigate any conversation and so the tendency of its motivation is to decrease only, except for the events when another agent engages it in a conversation.

\begin{figure}[h] 
\centering
 \includegraphics[width=1.0\columnwidth]{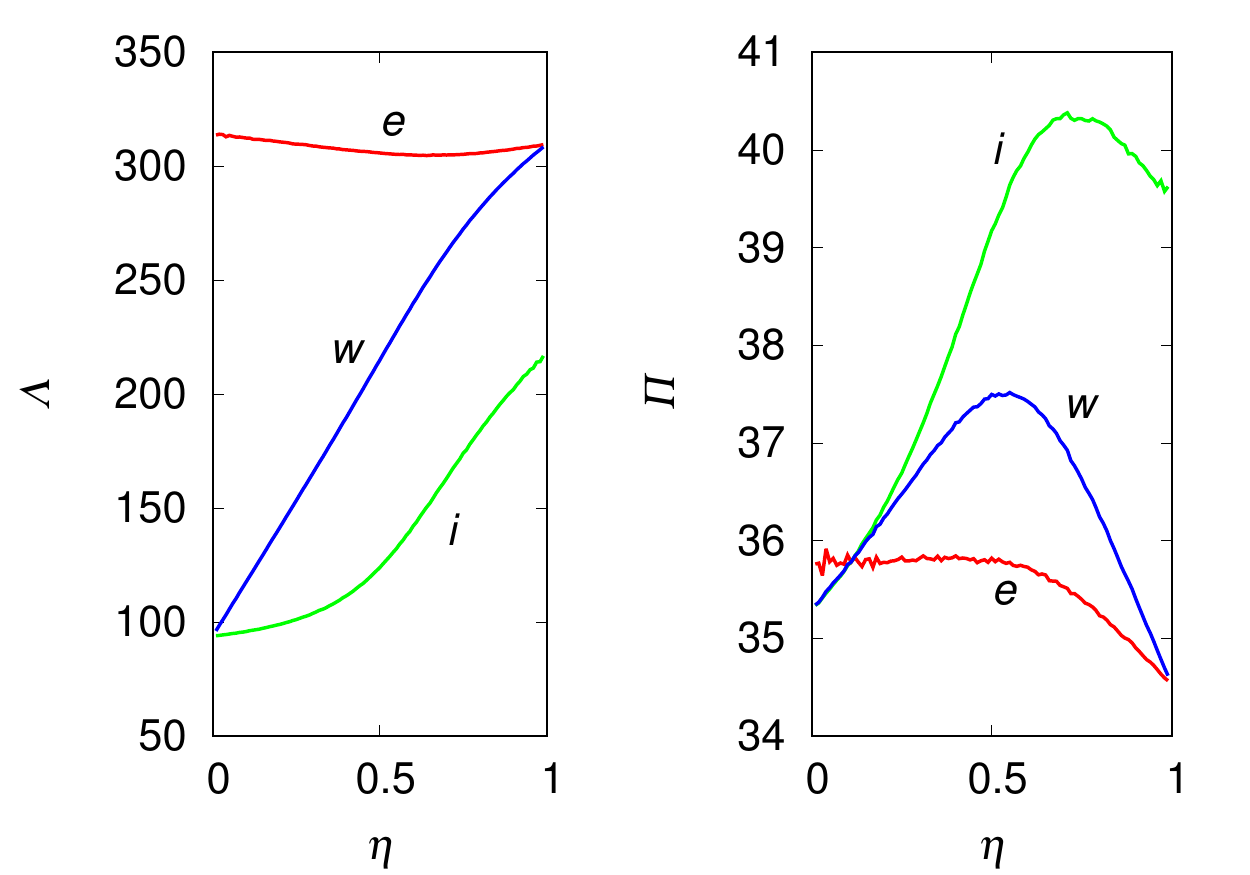}  
\caption{Effect of the fraction of extroverts on the mean motivation  (left panel) and on the mean cumulative productivity  (right panel)  at the end of the day  of the  subgroups of extroverts  ($e$) and  introverts  ($i$),  and of  the  whole group ($w$). The total number of agents is $N=100$,  the maximum duration of a conversation is $D=20$ and the mean number of contact attempts  is $q=2$.
 }  
\label{fig:2}  
\end{figure}

   Figure \ref{fig:2}  shows the influence of the fraction  of extroverts $\eta$ on the motivation and mean cumulative productivity of the agents at the end of the day. This figure reveals  some interesting features. First and foremost, the insertion of extroverts in  groups composed predominantly of introverts results in a significant increase of the mean productivity  of the introverts,  whereas the mean productivity  of the extroverts  is barely changed. Too many extroverts, however, cause a decrease on the productivity of both stereotypes, although they  considerably boost the motivation of the introverts. In addition, there is an optimum group composition ($\eta \approx 0.5$ for the parameters used in fig.\ \ref{fig:2}) that maximizes the total productivity of the group.
 
\begin{figure}[h] 
\centering
 \includegraphics[width=.9\columnwidth]{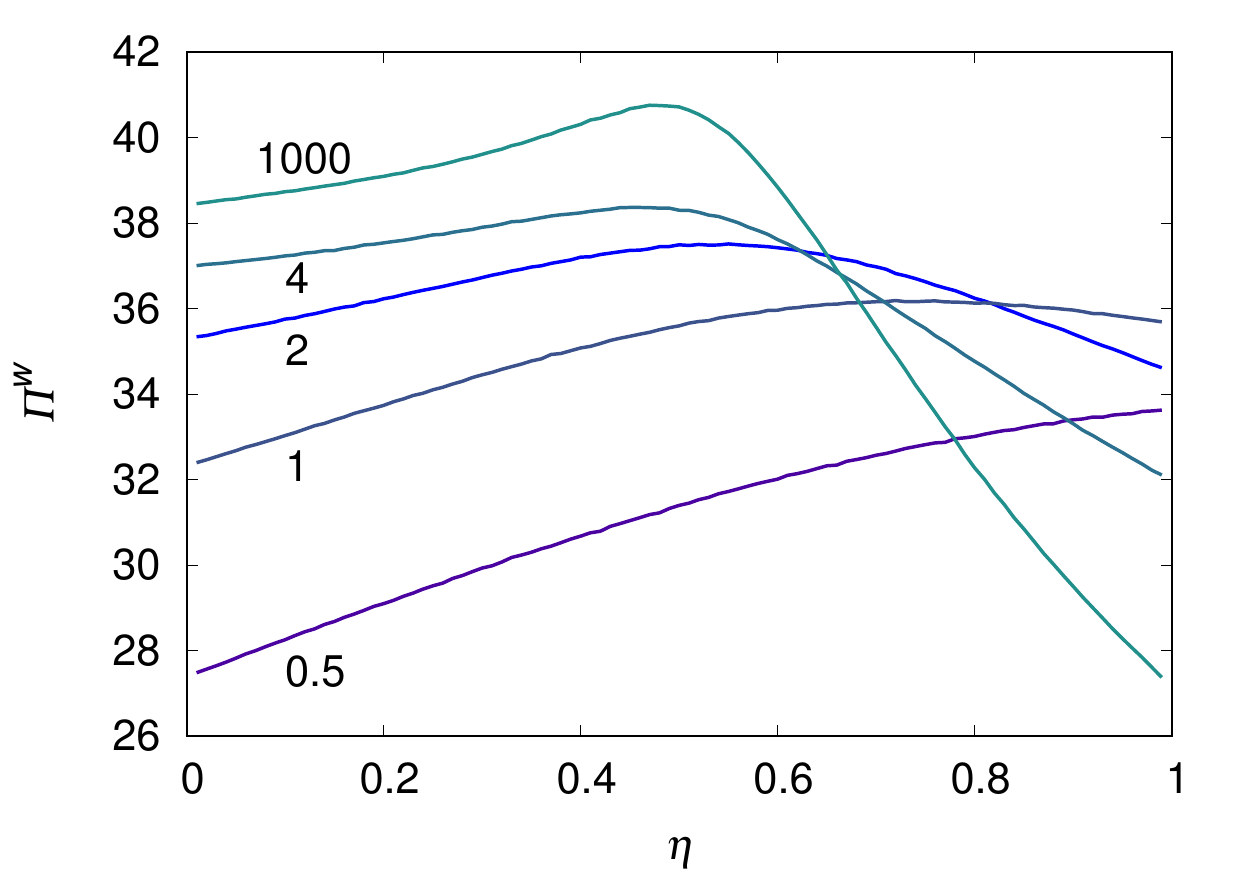}  
\caption{Mean cumulative productivity   of  the  whole group  at the end of the day  as function of the fraction  of extroverts  for 
 mean number of contact attempts   $q=0.5, 1, 2, 4$ and $1000$, as indicated. The total number of agents is $N=100$ and the maximum duration of a conversation is $D=20$.
 }  
\label{fig:3}  
\end{figure}

In fig.\ \ref{fig:3},  we examine the  influence of the mean number of contact attempts   $q $ on the optimum composition of the group. For instance, for  small  $q$, say $q=0.5$,  it is difficult to find conversation partners to increase the agents' motivations and so the highest total productivity is achieved by all-extroverts groups, since this stereotype seeks for partners more frequently than the introverts.  As $q$ increases,  the odds of finding a conversation partner increases as well, and so it becomes advantageous to insert a few introverts in the group. For very large $q$, so that the target agent attempts to contact all the other agents in the group, the  maximum total productivity occurs for $\eta \approx 0.47$.  For groups  composed predominantly of introverts ($\eta < 0.5$), increase of $q$ and $\eta$ lead always to higher total productivity. However, the scenario is more complicated for groups composed predominantly of extroverts. In fact, for $\eta > 0.5$ our results indicate that there is an optimum  value of the mean number of contact attempts  that maximizes the total productivity. For instance,   this optimum value is located between $q=1$ and $q=2$ for $\eta \approx 1$.

\begin{figure}[h] 
\centering
 \includegraphics[width=.9\columnwidth]{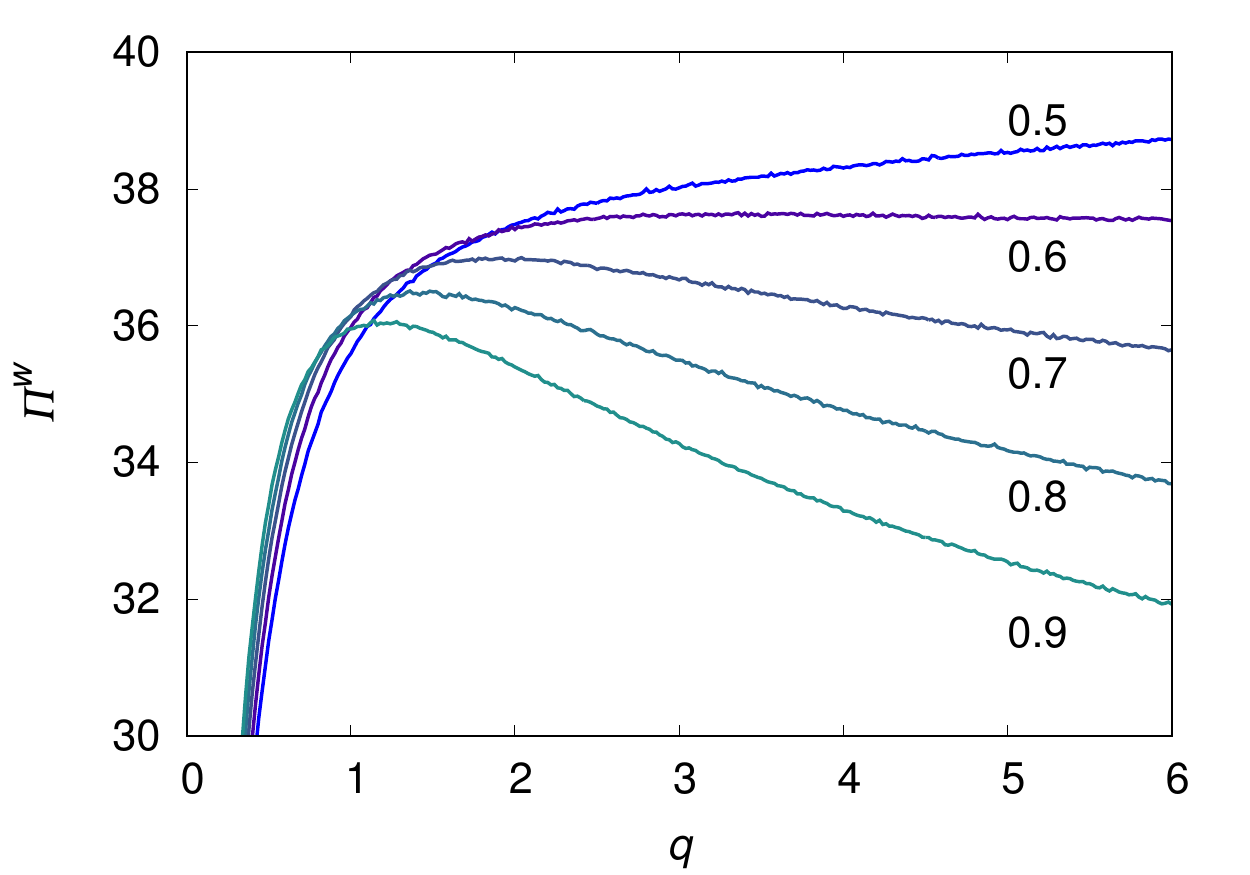}  
\caption{Effect of the mean number of  attempts to establish social interaction on the mean cumulative productivity  of  the  whole group   at the end of the day  for the  fraction of extroverts $\eta = 0.5, 0.6, 0.7, 0.8$ and $0.9$, as indicated. The total number of agents is $N=100$ and the maximum duration of a conversation is $D=20$.
 }  
\label{fig:4}  
\end{figure}

It is worthwhile to examine the effects of the parameter $q$ in more detail.  In the context of the 2020 coronavirus pandemic,  this is the leading parameter of the model since, at least for those adhering  to  social distancing norms,  the  mean number of  attempts to interact  socially  was  considerably curtailed  during the pandemic. Accordingly,  fig.\ \ref{fig:4}
 shows  that a moderate decrease of $q$ actually increases the productivity of the group for $\eta > 0.5$. For instance, for $\eta > 0.8$ the decrease from $q=6$ to $q=1$ results in a  productivity increase  of about $10\%$, which is in agreement with  the experiments of ref.\ \cite{Bloom_15}.  We note that in  both  our simulations  and those experiments the boost in  productivity  is  modest, though altogether unexpected.


\begin{figure}[h] 
\centering
 \includegraphics[width=.9\columnwidth]{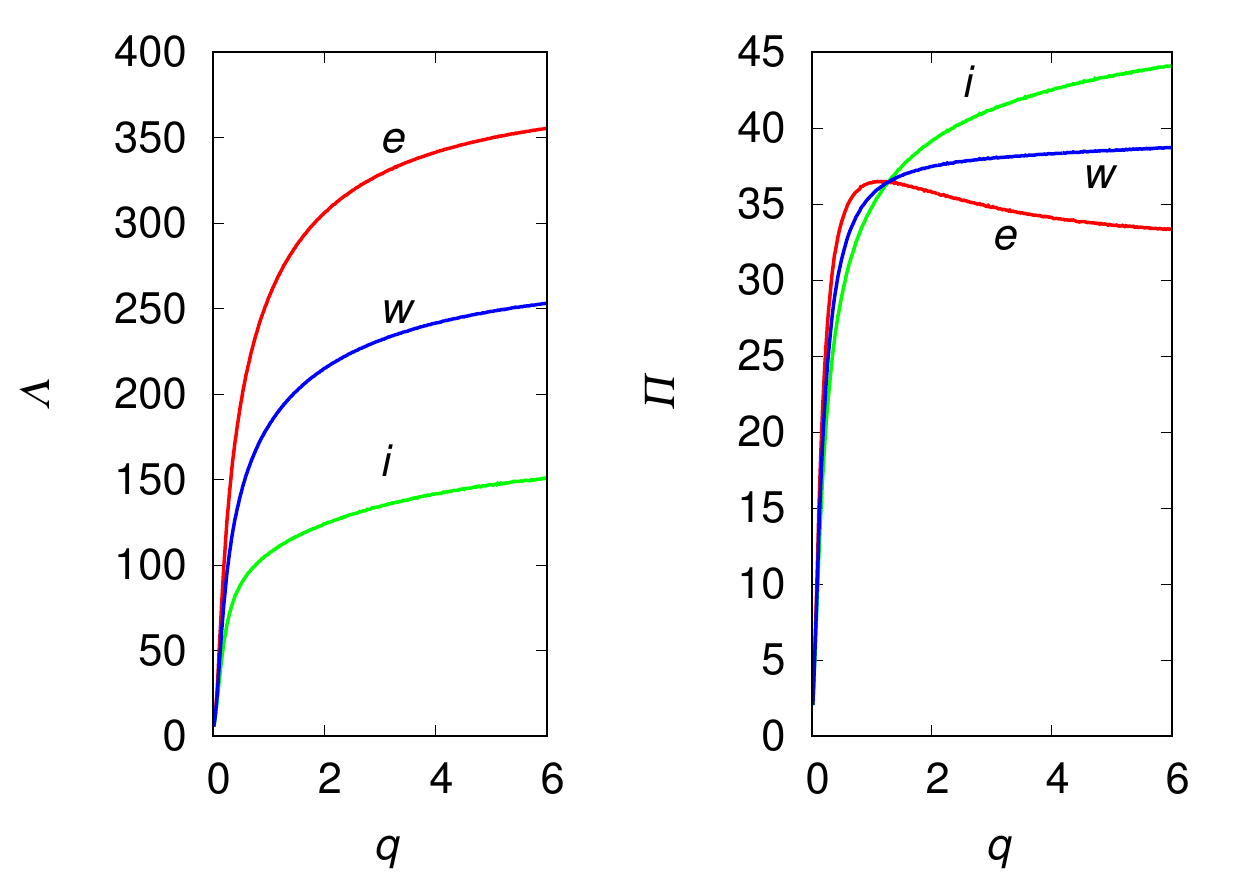}  
\caption{Effect of the mean number of  attempts to establish social interaction on the mean motivation  (left panel) and on the mean cumulative productivity  (right panel)  at the end of the day  of the  subgroups of extroverts  ($e$) and  introverts  ($i$),  and of  the  whole group ($w$). The total number of agents is $N=100$,  the maximum duration of a conversation is $D=20$ and the fraction of extroverts is $\eta=0.5$.
 }  
\label{fig:5}  
\end{figure}


  It is interesting to  look at how  the productivities of the two stereotypes within a same working group are  affected by the social restrictions. Figure  \ref{fig:5}  shows that, for $\eta = 0.5$,  a moderate decrease of $q$ reduces the motivation of both stereotypes as well as the mean productivity of the introverts,   but  increases the mean productivity of the extroverts. This finding may explain  sporadic personal reports of heightening productivity in the quarantine.
  Further decrease of $q$ leading to a scenario  where social interactions happen very rarely  results in a sharp drop of the mean productivity of both  stereotypes.
These results add to the message that  mathematical and computational  models  of the effects of the 2020 coronavirus pandemics and its palliative measures (e.g., social distancing) need to take into account the  distinct age and socioeconomic  segments of the population   \cite{Bellomo_20}. In fact, we argue here that  the  individuals' psychological traits  play an important role regarding work productivity and hence ought to be considered   on an assessment of the socio-economic implications of the pandemics.


A word is in order about the effects of the parameters $N$ and $D$ that determine the group size and the maximum duration of a conversation, respectively.  Reducing the group size has no effect on our results and our choice of fairly large groups  ($N=100$) is so as to smooth out  the variation of the fraction of extroverts $\eta$. However,  increase of the parameter $D$ has a strong effect on the productivity of the extroverts. For instance,  for long conversations, say $D=60$,  all-introverts groups yield the highest total productivity regardless of the number of contact attempts.  Our choice $D=20$, which  corresponds to conversations of average duration $\bar{d} = 10$, are meant to model  a ``water cooler talk'' scenario.

\textbf{Conclusion.} -- In the spirit of  sociophysics \cite{Perc_17,Perc_19} and  computational social science \cite{Lazer_09}, our  agent-based model builds on a variety of experimental findings of social psychology on the interplay between mood,  social interaction and productivity \cite{Stumm_18,Phillips_67,McIntyre_91,Isen_87,Ellis_88,Oswald_15} to advance an explanation for the mostly  anecdotal reports of increased productivity  in the present-day pandemics  scenario as well as for the more solid findings  that  productivity is increased  in some  remote work experiments \cite{Bloom_15}. These are puzzling findings because they challenge the established view that  communication among co-workers in the workplace increases  productivity either through peer pressure or knowledge spillover  \cite{Cornelissen_17,Kandel_92,Bandiera_10,Feldman_99}. The key ingredient to explain this puzzle is the observation that the population is   heterogeneous with respect to the individuals' propensities to engage in social interactions \cite{Feist_12} and that introverts and extroverts are affected differently by the social distancing and quarantine policies. 

Since the social psychology  studies  that support our model assumptions are  correlational,  they do not offer  specific quantitative relationships  between mood  and productivity  or between  mood and  propensity to initiate a conversation that are necessary to implement a simulation of the workplace scenario. Hence, to minimize spurious effects due to the, to some extent, arbitrary choices of the functional
relationships between those variables,  our model  assumes linear relationships  only. However, provided different choices accord with the correlational studies, we expect our conclusions to remain valid.  We stress that, rather than offering a precise quantitative characterization of the workplace dynamics, our model  offers a  proof of concept that quarantining and social distancing  may boost the productivity of extroverted  people, thus throwing light on  the paradox of productivity during quarantine.

\bigskip

\centerline{*~*~*}

The research of JFF was  supported in part 
 by Grant No.\  2020/03041-3, Fun\-da\-\c{c}\~ao de Amparo \`a Pesquisa do Estado de S\~ao Paulo 
(FAPESP) and  by Grant No.\ 305058/2017-7, Conselho Nacional de Desenvolvimento 
Cient\'{\i}\-fi\-co e Tecnol\'ogico (CNPq).




\end{document}